\documentclass[twocolumn,aps,prd,amsmath,amssymb,preprintnumbers,longbibliography]{revtex4-1}
\usepackage{amsmath} 
\usepackage{amsfonts} 
\usepackage{amssymb}
\usepackage{bbm}
\usepackage{epsfig}
\usepackage{graphics}
\usepackage{graphicx}
\usepackage{titlesec}
\usepackage{mathtools}
\usepackage{environ}

\makeatletter
\renewcommand*\env@matrix[1][c]{\hskip -\arraycolsep
  \let\@ifnextchar\new@ifnextchar
  \array{*\c@MaxMatrixCols #1}}
\makeatother

\newcommand{\be}{\begin{equation}}
\newcommand{\ee}{\end{equation}}
\newcommand{\ba}{\begin{eqnarray}}
\newcommand{\ea}{\end{eqnarray}}

\titleformat{\subsection}[block]{\normalfont\bfseries}{\thesubsection.}{1ex}{}
\titlespacing{\subsection}{0pt}{10pt}{1pt}[0pt]
\titleformat*{\section}{\large\bfseries}
\renewcommand{\thesubsection}{\arabic{subsection}}

\DeclareMathOperator{\tr}{tr}
\DeclareMathOperator{\Det}{Det}
\DeclareMathOperator{\Str}{Str}
\newcommand{\transp}{\mathrm{T}}

\newenvironment{alignedeqn}{\begin{equation}\begin{aligned}}{\end{aligned}\end{equation}\ignorespacesafterend}

\usepackage[colorlinks=true]{hyperref}
\hypersetup{urlcolor=black,linkcolor=black,citecolor=black}

\begin{document}

\title[ ]{Gauge-invariant flow equation}

\author{C. Wetterich}
\affiliation{Institut  f\"ur Theoretische Physik\\
Universit\"at Heidelberg\\
Philosophenweg 16, D-69120 Heidelberg}

\begin{abstract}
    We propose a closed gauge-invariant functional flow equation for Yang-Mills theories and quantum gravity that only involves one macroscopic gauge field or metric. It is based on a projection on physical and gauge fluctuations. Deriving this equation from a functional integral we employ the freedom in the precise choice of the macroscopic field and the effective average action in order to realize a closed and simple form of the flow equation.
\end{abstract}

\maketitle

\section{Introduction}

Functional flow equations permit to interpolate continuously from the microscopic or classical action to the macroscopic or quantum effective action. For Yang Mills theories and quantum gravity local gauge symmetries play a central role. A functional renormalization approach to such theories should keep carefully track of gauge symmetries and resulting restrictions on the general form of the effective action.

The goal is to realize a gauge-invariant effective action $\bar{\Gamma}(\bar g)$ for a single metric $g_{\mu\nu}$ in gravity, or a single gauge potential $A_\mu = g_\mu$ for electromagnetism, once all fluctuations are taken into account. (Quantum) field equations are then obtained directly as $\partial \bar\Gamma/\partial \bar g = K$, with $K$ an appropriate conserved source. These field equations are the basis for the ``classical'' field theories of gravity and electromagnetism, which are well tested by many precision observations. A similar gauge-invariant effective action will be formulated for Yang-Mills theories. Physical correlations or Green's functions are obtained by inverting the second functional derivative of $\bar\Gamma$ in the space of physical fluctuations.

At the present stage, the formulation of functional flow equations for gauge theories has to deal with the problem that regularization in continuum field theories typically breaks the gauge symmetry, necessitating gauge fixing. Furthermore, quadratic infrared cutoff terms are usually not compatible with the gauge symmetry. Exact flow equations for the effective average action of gauge theories have been formulated in the background field formalism \cite{RW1,RWA,RWB,RWC}. This formalism has been extended to quantum gravity \cite{MR1}. These flow equations involve, however, two independent fields. The first is the expectation value of the microscopic or fluctuating field $g^\prime$, over which the functional integral is performed,
\begin{equation}
    g
    = \langle g^\prime\rangle,
\end{equation}
while the second ``background field'' $\bar{g}$ is used to formulate covariant derivatives for the gauge fixing and infrared cutoff. The effective action is only invariant under simultaneous transformations of $g$ and $\bar g$. 

Alternatively, one  may omit the background field, which amounts to setting $\bar g=0$ in the background field formalism. The effective action is no longer gauge invariant. Rather sophisticated approximation schemes \cite{CLPR,CKPR} are needed in order to cope with the many terms contributing already in low orders of the gauge field. It has been proposed to maintain gauge symmetry by the use of rather complex gauge invariant regularizations \cite{TM2,MO,MOR,FM} involving additional fields. Our present approach is more modest. Technically, it shows analogies to background gauge fixing in a particular ``physical'' gauge. We obtain, however, a gauge invariant effective action depending only on one macroscopic gauge field. This is achieved by employing the macroscopic field for the formulation of the gauge fixing and infrared regulator term. No separate background field is introduced. At the end, we obtain indeed a quantum effective action that is gauge invariant and depends on a single metric or gauge field. This can be used as the basis for general relativity and Maxwells equations, including corrections to these equations generated by quantum fluctuations.

In the usual ``background field formalism'' $\bar{g}$ is considered as fixed. We propose here to replace the fixed background field by a macroscopic field $\bar{g}(g)$, with a relation to $g$ that is, in principle, computable. The macroscopic field $\bar{g}$ is the argument of the gauge-invariant effective action $\bar\Gamma(\bar{g})$ which only depends now on one field. The metric or gauge field in the field equations is identified with $\bar{g}$. Also the flow equations describe the scale dependence of the effective action at fixed $\bar{g}$. Thus $\bar{g}$ is the relevant field for all macroscopic considerations. (We keep here the notation $\bar{g}$ for comparison with the background field formalism -- the bar may be dropped at later stages.) The choice of the relation between $\bar{g}$ and $g=\langle g^\prime\rangle$ is such that a closed gauge invariant flow equation can be formulated for $\bar\Gamma(\bar{g})$. The precise relation between $ g$ and $\bar{g}$ is of secondary importance.

Approximative solutions (truncations) of previous versions of exact flow equations for gauge theories have been successfully used to understand various phenomena. Superconductivity or the abelian Higgs model has been investigated in various dimensions \cite{RWA,RWC,BFLLW,BLLW2}. Increasingly sophisticated truncations in quantum chromodynamics (QCD) provide for an increasingly complete analytical understanding \cite{G2,LP1,PA1}. Functional renormalization has addressed the running of the gauge coupling in various dimensions \cite{RW1,RWB,EHW,G1,G4}. Applied to thermal equilibrium, with an effective non-perturbative (``confinement'') scale increasing with temperature, it has been advocated that non-perturbative strong interaction effects should be visible in the quark gluon plasma even at high temperature \cite{RW1,BG}. (This qualitative finding has been made quantitative by computations of thermodynamic quantities in lattice gauge theories, or by the experimental observation of strong interaction properties in heavy ion collisions at high effective temperature.) Detailed studies of the flow in the non-perturbative regime have addressed the issues of the heavy quark potential \cite{CWEQ,CWGF,BW}, gluon condensation \cite{RW2,EGP}, the gluon propagator \cite{CWGF,BW,CFM} and confinement \cite{BGP}. The comparison of flow equation results in Landau gauge with lattice simulations \cite{CFM,SIM} and Schwinger-Dyson equations \cite{FP1,FP2,CFM,MPS} have added confidence in the reliability of results for QCD. For the electroweak interactions the crossover character of the high-temperature transition has first been advocated based on the effective three-dimensional running of couplings \cite{RWB}.

In quantum gravity the non-perturbative flow equations for the effective average action have permitted to address the asymptotic safety scenario \cite{Wei} in four dimensions \cite{MR1}. The corresponding ultraviolet (UV) fixed point of the flow \cite{MR1,Sou} has been seen to persist for rather extended truncations \cite{Dou:1997fg,Reuter:2001ag,Litim:2003vp,Codello:2006in,Machado:2007ea,Codello:2008vh,Fischer:2006fz,Benedetti:2009rx,Eichhorn:2010tb,Donkin:2012ud,Christiansen:2012rx,Rechenberger:2012dt,Dietz:2012ic,Codello:2013fpa,Falls:2013bv,Benedetti:2013jk,Christiansen:2014raa,Christiansen:2015rva,Dietz:2015owa,Demmel:2015oqa,Falls:2015qga,Gies:2015tca,Gies:2016con}. Within dilaton quantum gravity a similar UV-fixed point can be related directly to inflationary cosmology \cite{H1,H2}. Despite these many striking successful applications of functional flow equations for gauge theories, further progress is partially hindered by the proliferation of the number of invariants in the absence of a realization of gauge symmetry for a single gauge field. While the conceptual setting and the exactness of the flow equation is not in doubt, the absence of gauge symmetry or the presence of two gauge fields in the background field formalism makes it hard to derive series of truncations that do not rapidly become very complex.

In the background field formalism the flowing action or effective average action $\Gamma(g,\bar{g})$ is gauge invariant if $g = \langle g^\prime\rangle$ and $\bar{g}$ are transformed simultaneously. In contrast, gauge invariance is broken if only $g^\prime$ and $g$ are transformed while $\bar{g}$ is held fixed. A gauge-invariant effective action involving only one field, $\Gamma(\bar{g}) = \Gamma(\bar{g},\bar{g})$, can be formed if $g$ is identified with $\bar{g}$. This object is in the center of many studies in the past. The exact flow equation for $\Gamma(\bar{g})$ involves, however, the exact propagator which is encoded in $\Gamma(g,\bar{g})$ \cite{RW1}. Indeed, the inverse propagator is given by the second functional derivative of $\Gamma(g,\bar{g})$ with respect to $g$, taken at fixed $\bar{g}$. It is not directly related to the second functional derivative of $\Gamma(\bar{g})$. What is needed is an estimate of the shape and influence of
\begin{equation}\label{eqn:I1}
	\Delta\Gamma(g,\bar{g})
	= \Gamma(g,\bar{g}) - \Gamma(\bar{g},\bar{g}).
\end{equation}
Many practical computations assume that $\Delta\Gamma$ can be sufficiently well described by a simple gauge fixing term. A reliable estimate of the effects from $\Delta\Gamma$ beyond such a simple ansatz is perhaps the most important present source of uncertainty and error in the functional renormalization group approach to gauge theories and quantum gravity.

The functional form of $\Delta\Gamma(g,\bar{g})$ obeys various constraints which guarantee that there are no physical degrees of freedom beyond the ones contained in a single gauge field. An exact ``background field identity'' \cite{RW1} yields a one loop type exact equation for the dependence of the effective action on the background field at fixed $g$, i.e. $\partial\Gamma(g,\bar{g})/\partial\bar{g}|_g$. In the absence of an infrared cutoff ($k = 0$) gauge theories with gauge fixing obey exact Ward or Slavnov-Taylor identities. For an infrared cutoff scale $k \neq 0$ those are violated by the infrared cutoff. Exact modified Ward identities \cite{BEC,EL,BAM,AM} do now account for this violation of the usual Ward identities by effects of the infrared cutoff. The modified Ward identities can be obtained from the background field identity \cite{FW,FLP,LP3} - both sets of identities express the same physics, namely the absence of unphysical propagating degrees of freedom. Already in a perturbative setting these identities are cumbersome to handle, however, and there is very little experience how to implement these often rather complex identities in a non-perturbative situation.

Alternatively, one may directly compute the flow of $\Delta\Gamma$ by an exact flow equation \cite{MR,MRS,BEM}. Since $\Delta \Gamma$ involves $h = g - \bar{g}$, and $h$ transforms homogeneously as a tensor under simultaneous gauge transformations of $g$ and $\bar{g}$, possible truncations get quickly rather involved. There are many invariants that can be constructed from the tensor $h$. Keeping the dependence on $h$ and $\bar{g}$ unconstrained results in new relevant operators in the flow. The values of their coefficients have to be tuned in order to maintain consistency with the background field identity or the modified Ward identities. It is precisely these identities that are responsible for the absence of physical additional relevant couplings in the two-field formalism. Since the solutions of the identities are only poorly known, it is often difficult to achieve the physical tuning in a given truncation, resulting in potentially large errors.

In this note we propose to avoid the problems with $\Delta\Gamma(g,\bar{g})$ altogether by the definition of a flowing gauge-invariant effective action $\bar\Gamma(\bar{g})$ which admits a closed flow equation. This means that the flow generator $\zeta_k$ (r.h.s. of the flow equation) can be expressed as a functional of $\bar\Gamma(\bar{g})$, typically involving the second functional derivative $\bar\Gamma^{(2)}(\bar{g})$. Since $\bar\Gamma^{(2)}(\bar{g})$ has zero modes due to gauge symmetry, the propagator for the physical modes has to be found by inversion on a suitably projected subspace. Our formalism shows some analogies with ``geometric flows'' \cite{P5,Donkin:2012ud,DSZ}.

In practice, computations with the proposed gauge-invariant flow equation are rather similar to computations in the background field formalism for a specific physical gauge. The flow equation retains its one-loop form. Due to the dependence of the cutoff function  $R_k$ on $\bar{g}$ via covariant derivatives, the flow of derivatives of $\bar\Gamma$ involves additional diagrams $\sim \partial R_k/\partial \bar{g}$. For $k \neq 0$ the inverse propagator obtained from the second functional derivative of $\bar\Gamma[\bar{g}]$ does not equal the connected two-point function for the microscopic fluctuations, however.

The paper is organized as follows: In sect. \ref{Flow of gauge-invariant effective action} we describe the proposed gauge invariant flow equation and establish its gauge invariance. Sect. \ref{Flow equation from functional integral} derives the flow equation from a functional integral. In contrast to the usual formulation the partition function depends on the macroscopic gauge field $\bar g$, rendering the definition of the effective action an integro-differential functional equation. This issue and the consequences for the flow equation are discussed in sect. \ref{Macroscopic field}. In sect. \ref{Projection on physical fluctuations} we describe the projection on the physical fluctuations and the notion of a physical gauge fixing that acts solely on the gauge fluctuations. The optimal choice of the macroscopic gauge field is discussed in sects. \ref{Choice of macroscopic field} and \ref{Optimal macroscopic field}. Only this optimal choice permits a simple closed form of the flow equation. The short sect. \ref{Quantum field equation} addresses the quantum field equations that are the basis for classical field theory. We discuss our results in sect. \ref{Discussion}.

\section{Flow of gauge-invariant effective action}
\label{Flow of gauge-invariant effective action}

We start our discussion by proposing a flow equation for a gauge invariant functional $\bar\Gamma_k(\bar g)$ that depends only on the macroscopic gauge field $\bar g$. This equation for the evolution with the infrared scale $k$ is closed, such that specifying the ``initial condition'' at some ultraviolet value of $k$ permits, in principle, to extract the gauge invariant effective action as the solution of the flow equation at $k=0, ~\bar \Gamma(\bar g)=\bar\Gamma_{k=0}(\bar g)$. The first derivative of $\bar\Gamma$ defines the field equation and the second the inverse propagator. If the usual relation between the two-point correlation of fluctuations and the propagator holds, $\bar\Gamma(\bar g)$ is sufficient to compute the quantities of interest for gravity, electromagnetism or Yang-Mills theories. In a second step we will relate this flow equation to a functional integral and discuss if it is exact or should be considered as an approximation. 

The flow equation involves the exact propagator for the physical fluctuations. It is therefore important to separate the physical fluctuations and the gauge fluctuations by appropriate projections. Only on the projected subspace for physical fluctuations the second functional derivative of $\bar\Gamma[\bar g]$ is invertible and related to the propagator of the physical fluctuations. The gauge invariant effective action $\bar\Gamma[\bar g]$ contains no gauge fixing term. In the space all fluctuations the second functional derivative of $\bar\Gamma[\bar g]$ has zero modes and is not invertible. The projections avoid the problems with the zero modes.

The proposed flow equation for the gauge-invariant effective average action $\bar\Gamma_k(\bar{g})$,
\begin{equation}\label{eqn:A}
	k\partial_k\bar\Gamma_k
	= \zeta_k
	= \pi_k + \delta_k - \epsilon_k,
\end{equation}
involves a contribution $\pi_k$ from the physical fluctuations as well as ``measure contributions'' $\delta_k$ and $\epsilon_k$. Only $\pi_k$ depends on $\bar\Gamma$, according to
\begin{equation}\label{eqn:B}
	\pi_k
	= \frac{1}{2} \tr (k\partial_k\bar{R}_PG_P),
\end{equation}
with $\bar{R}_P$ the infrared regulator for the physical fluctuations which depends on the normalization scale $k$. The trace $\tr$ contains a momentum integration such that eq. \eqref{eqn:B} takes a one loop form, as well as a suitable trace over Lorentz- and internal indices of the gauge fields or the metric. All indices, including position or momentum labels, are collectively denoted by $i$.

The propagator matrix $G_P$ for the physical fluctuations obeys the projection properties $PG_P = G_PP^\transp = G_P$, with projector $P = P^2$. The same holds for $\bar{R}_P = P^\transp\bar{R}_P = \bar{R}_PP$. The projector is defined such that it projects on the physical fluctuations. It annihilates the infinitesimal gauge variations of the variable $\bar{g}$ according to
\begin{equation}\label{eqn:C}
	\delta_\xi\bar{g}
	= (1 - P)\delta_\xi\bar{g}~,~P\delta_\xi\bar g=0.
\end{equation}
(Here $\bar{g}$ is considered as a vector with index $i$, e.g. $\bar{g}_i = A^z_\mu(x)$ or similar. Correspondingly, $P,\bar G_P$ and $\bar{R}_P$ are matrices.)

The projector is a given functional of the macroscopic field $\bar{g}$, determined uniquely by the action of gauge transformations on $\bar{g}$. For the example of non-abelian gauge theories the projector reads \cite{RW1}
\begin{equation}\label{eqn:N}
	P_\mu{^\nu}
	= \delta^\nu_\mu - \bar P_\mu{^\nu},
	\qquad
	\bar P_\mu{^\nu}
	= D_\mu D^{-2} D^\nu,
\end{equation}
with $D_\mu$ the covariant derivative in the adjoint representation involving the macroscopic field $\bar{g} = \bar A_\mu$, and $D^2 = D_\rho D^\rho$. For quantum gravity the projector again involves covariant derivatives and therefore depends on $\bar g$. It is discussed in ref. \cite{CWQC}.

We define the projected second derivative of $\bar\Gamma$ by
\begin{equation}\label{eqn:D}
	\bar\Gamma^{(2)}_P
	= P^\transp\bar\Gamma^{(2)}P,
	\qquad
	\bar\Gamma^{(2)ij}
	= \frac{\partial^2\bar\Gamma}{\partial\bar{g}_i\partial\bar{g}_j}.
\end{equation}
The quantity $\bar\Gamma^{(2)}_P + \bar{R}_P$ is invertible on the space of physical fluctuations. We define the propagator $G_P$ as the inverse of $\bar\Gamma_P^{(2)}$ in the projected space of physical fluctuations,
\begin{equation}\label{eqn:E}
	\left(\bar\Gamma^{(2)}_P + \bar{R}_P\right)G_P
	= P^\transp.
\end{equation}
Thus $G_P$ is computable in terms of $\bar\Gamma$, and the flow equation \eqref{eqn:A}, \eqref{eqn:B} is closed. Without the projection on physical fluctuations the matrix $\bar\Gamma^{(2)}$ is not invertible due to the presence of gauge modes and the gauge invariant construction of the propagator \eqref{eqn:E} would not be possible.

The measure contributions are fixed functions of suitable differentiable operators. In a gauge fixed version they account for the contribution from gauge fluctuations and the Faddeev Popov determinant for a ``physical gauge fixing''. For Yang-Mills theories one has 
\begin{equation}\label{XAB}
 \delta_k-\epsilon_k=-\frac12 \text{tr}\Big\{k\partial_kR_{gf}({\cal D}_S)\big({\cal D}_S+R_{gf}({\cal D}_S)\big)^{-1}\Big\},
\end{equation}
with ${\cal D}_S=-D^\mu D_\mu$ and $R_{gf}$ a suitable infrared cutoff function. The covariant derivative $D_\mu$ is formed with the macroscopic gauge field $\bar g$, such that $\delta_k-\epsilon_k$ is gauge invariant. This part of the flow can be computed independently of the precise form of $\bar\Gamma_k$ and its possible truncations. The main differences of the proposed flow equation \eqref{eqn:A} as compared to the exact flow equation for scalars or fermions is the projection on physical fluctuations and the related presence of a measure contribution.

A gauge-invariant effective action obeys
\begin{equation}\label{eqn:F}
	\frac{\partial\bar\Gamma}{\partial\bar{g}}
	= \frac{\partial\bar\Gamma}{\partial\bar{g}}P,
\end{equation}
such that
\begin{equation}\label{eqn:G}
	\delta_\xi\bar\Gamma
	= \frac{\partial\bar\Gamma}{\partial\bar{g}}P\delta_\xi\bar{g}
	= 0.
\end{equation}
We want to show that $\pi_k$ is gauge invariant, e.g. $\partial \pi_k/\partial\bar g=(\partial\pi_k/\partial\bar g)P$. Gauge invariance is then preserved by the flow. If one starts with a gauge invariant functional $\Gamma_\Lambda$ at some scale $k=\Lambda$, the effective average action $\Gamma_k$ remains gauge invariant for all $k<\Lambda$. 

For this purpose we first note that $\bar\Gamma^{(2)}$ in eq. \eqref{eqn:D} transforms homogeneously as a symmetric tensor (e.g. rank four for gravity, rank two and adjoint in internal space for Yang-Mills theories). The projectors involve covariant derivatives and transform as tensors as well. This implies that $\bar\Gamma^{(2)}_P$ in eq. \eqref{eqn:D} transforms as a tensor. We choose $\bar R_P$ to have the same tensor transformation as $\bar\Gamma^{(2)}_P$. This is straightforward if one uses covariant derivatives depending on $\bar g$ for its construction. From eq. \eqref{eqn:E} one infers that the projected propagator $G_P$ transforms as a tensor as well, consistent with a correlation function of field fluctuations. Finally, the derivative $k\partial_k R_k$ also transforms as a tensor, and the r.h.s of eq. \eqref{eqn:B} is therefore gauge invariant. 

More in detail, one can convince oneself that the gauge variation of the projector does not contribute. Formally, one may write
\begin{equation}\label{eqn:J}
	\bar{R}_P
	= P^\transp\bar{R}P,
	\qquad
	G_P
	= PGP^\transp,
\end{equation}
and gauge variation of $\pi_k$ could receive contributions from
\begin{equation}\label{eqn:K}
	\delta_\xi P
	= \Delta_P.
\end{equation}
From the projector properties $P = P^2 = P^3$ one derives for $\Delta_P$ the relation
\begin{equation}\label{eqn:L}
	\Delta_P
	= \bar P\Delta_P P + P\Delta_P \bar P,
	\qquad
	\bar P
	= 1 - P.
\end{equation}
As a consequence of the structure of \eqref{eqn:B} the factor $\bar P$ in $\Delta_P$ always gets multiplied by $P$, and with $\bar PP = 0$ we conclude that the gauge variation $\Delta_P$ does not contribute.

For Yang-Mills theories the relevant differential operators are explicitly known. Both $\delta_k$ and $\epsilon_k$ are traces of functions of $D^2$. Similarly, the IR cutoff for the physical fluctuations $\bar{R}_P$ is a function of the covariant operator
\begin{equation}\label{eqn:O}
	(\mathcal{D})_\mu{^\nu}
	= -D^2\delta^\nu_\mu + D_\mu D^\nu + 2 i {F}{_\mu^\nu},
\end{equation}
where ${F}{_\mu^\nu}$ is the contraction of the field strength $F_\mu^{z\nu}$ with the generator $T_z$ in the adjoint representation. For $D^\nu \, F^{\mu\nu} = 0$ it obeys $P \mathcal{D} = \mathcal{D} P = \mathcal{D}$. For gravity the explicit form of projected differential operators is not known for arbitrary background fields. For a large class of interesting backgrounds the classification of physical fluctuations remains nevertheless rather simple \cite{CWQC}.

More generally, explicit knowledge or use of the non-local projectors is often not needed for practical computations. Typically, $\bar\Gamma^{(2)}$ automatically satisfies the projection property in eq. \eqref{eqn:D}, and similar for $\bar{R}$. Solutions for $G_P$ obeying eq. \eqref{eqn:E} can be found without the need of explicit knowledge of $P$. For example, $\bar\Gamma_P^{(2)}$ and $G_P$ are often both proportional to $P = P^\transp$, such that only $\tr P$ is needed for eq. \eqref{eqn:B}. The trace is typically known, given simply by the number of physical degrees of freedom. Furthermore, the projection onto $G_P$ can also be realized by adding ``by hand'' a ``physical gauge fixing term'' with infinite coefficient $1/\alpha$. Inversion of the second functional derivative of the effective action in presence of this physical gauge fixing projects onto $G_P$, with all other components of the inverse of $\Gamma^{(2)}$ vanishing $\sim \alpha$. 

If the solution of the proposed gauge-invariant flow equation belongs to the same universality class as usual gauge theories, explicit knowledge of the microscopic formulation of the gauge-invariant effective average action $\bar\Gamma$ is actually not needed. We could use the flow equation \eqref{eqn:A} for an ``ERGE-regularization'' of a non-abelian gauge theory or quantum gravity. A derivation of the flow equation \eqref{eqn:A} from a functional integral is, nevertheless, relevant for several issues. For a functional integral the conditions for the description of a unitary quantum field theory (e.g. Osterwalder-Schrader positivity) are well established. Furthermore, the functional integral formulation of $\bar\Gamma_k(\bar g)$ makes the connection of the proposed flow equation to other formulations of gauge theories more apparent. This will also shed light on the choice of the measure contributions $\delta_k-\epsilon_k$. For a given microscopic functional integral formulation we can also address the question if eq. \eqref{eqn:A} is exact or if it is some type of approximation. If exact, the functional integral can be viewed as a formal solution of the differential flow equation. The remainder of this note will discuss the functional integral representation of $\bar\Gamma_k(\bar g)$. 

\section{Flow equation from functional \newline integral}
\label{Flow equation from functional integral}

We will next discuss the emergence of the flow equation from a microscopic functional integral formulation \cite{CWFE}. We investigate here a continuum formulation with gauge fixing and do not address the relation of this continuum formulation to possible discrete (lattice) gauge invariant formulations. Starting from a particular physical background gauge fixing we derive the flow equation \eqref{eqn:A} in two steps. For the first step we keep an arbitrary field $\bar{g}$ independent of $g = \langle g^\prime\rangle$. This closely follows ref. \cite{RW1}. In the second step we choose a suitable relation between $\bar{g}$ and $g$ which  relates the macroscopic field $\bar{g}$ to a nonlinear function of the expectation value of the microscopic field $g'$. 

Our starting point is the usual functional integral for the partition function in presence of a background field $\bar{g}$,
\begin{alignedeqn}\label{eqn:B1}
	&\begin{aligned}Z(L,\bar{g})
	= \int \mathcal{D} g^\prime M_k(g^\prime,\bar{g}) &\exp\bigl\{-S(g^\prime) - S_\text{gf}(g^\prime,\bar{g})\\
	&- \Delta S_k(g^\prime,\bar{g}) + L^\transp g^\prime\bigr\},
	\end{aligned}\\
	&W(L,\bar{g})
	= \ln Z(L,\bar{g}).
\end{alignedeqn}
The microscopic field $g^\prime$ may be considered as a generalized vector $g^\prime_i$, with indices $i$ including space or momentum labels, Lorentz indices $\mu$, indices for the representations of the gauge group $z$, and labels for different species. For the example of a pure Yang-Mills gauge theory $g^\prime$ stands for the gauge fields $A^{\prime z}_\mu(x)$, while for quantum gravity it denotes the microscopic metric $g_{\mu\nu}^\prime(x)$. The corresponding sources are denoted by $L, L^\transp g^\prime = L^i g^\prime_i$.

The microscopic action is given by $S(g^\prime)$, while the background field appears in the gauge fixing term $S_\text{gf}(g^\prime,\bar{g})$ and the infrared cutoff term $\Delta S_k$. The factor
\begin{equation}\label{eqn:B1A}
	M_k
	= M(g^\prime,\bar{g}) \, E_k(\bar{g})
\end{equation}
contains the Faddeev-Popov determinant $M$ and an associated regulator $E_k$ \cite{RW1}, with
\begin{equation}\label{eqn:B1B}
	\epsilon_k(\bar{g})
	= \tr\big \{\ln k\partial_k E_k (\bar{g})\big\}.
\end{equation}
This defines the measure contribution $\epsilon_k$ in \eqref{eqn:A}. The infrared cutoff term $\Delta S_k(g^\prime,\bar{g})$ vanishes for $k = 0$. In this limit \eqref{eqn:B1} becomes the standard setting for a gauge theory with background gauge fixing. For $k \to \infty$ the infrared cutoff should remove the fluctuation contributions such that $\bar\Gamma_{k \to \infty}(\bar{g}) = S(\bar{g})$. We discuss more details below.

As usual, one may define an effective action by a Legendre transform at fixed $\bar{g}$
\begin{equation}\label{eqn:B2}
	\tilde\Gamma(g,\bar{g})
	= -W(L,\bar{g}) + L^\transp g,
\end{equation}
with $g$ and $L$ related by
\begin{equation}\label{eqn:B3}
	g
	= \frac{\partial W(L,\bar{g})}{\partial L}
	= \langle g^\prime\rangle,
	\qquad
	L
	= \frac{\partial \tilde\Gamma(g,\bar{g})}{\partial g}.
\end{equation}
For suitable choices of $S_\text{gf}$ and $\Delta S_k$ the effective action \eqref{eqn:B2} is invariant under simultaneous gauge transformations of $g$ and $\bar{g}$. This is usually called ``background gauge symmetry'', while ``microscopic gauge transformations'' only transform $g'$, and therefore $g$, leaving $\bar g$ fixed. The effective action $\tilde \Gamma$ is not invariant under microscopic gauge transformations. 

As the key idea of this note we employ here a macroscopic field $\bar{g}(g)$ which depends on the expectation value $g$, rather than a fixed value. Since $\bar g(g)$ and $g$ do not transform independently, a distinction between background gauge transformations and microscopic gauge transformations is no longer possible. All fields $g',g$ and $\bar g(g)$ transform under a single gauge transformation. Inserting $\bar{g} (g)$ in $\tilde\Gamma(g,\bar{g})$ yields an effective action that only depends on one field. As independent variable we choose the macroscopic field $\bar g$, with $g$ expressed in terms of $\bar{g}$ by inverting $\bar{g}(g)$. The use of $\bar{g}(g)$ in the gauge fixing and infrared cutoff terms transmutes the defining equation for $W$ into an integro-differential equation. Now $\partial W/\partial L$ appears in the gauge-fixing term, $\Delta S_k$ and $M_k$ through $\bar{g}(g) = \bar{g}(\partial W/\partial L)$. We will see, however, that there is no need to solve this integro-differential equation explicitly. We emphasize that the definition of $\tilde\Gamma(g,\bar{g})$ is the Legendre transform of $W(L,\bar{g})$ at \textit{fixed} $\bar{g}$. One could define a different object as the Legendre transform of $W(L) = W\big (L,\bar{g}(L)\big)$. This is not what we use here.

An exact flow equation for $\tilde \Gamma(g,\bar{g})$ can be derived \cite{RW1} by varying the infrared cutoff term in eq. \eqref{eqn:B1} which should be at most quadratic in the microscopic field $g^\prime$,
\begin{equation}\label{eqn:B4}
	\Delta S_k(g^\prime,\bar{g})
	= \frac{1}{2}(g^\prime - \bar{g})^\transp R_k(\bar{g})(g^\prime - \bar{g}).
\end{equation}
The cutoff function $R_k$ is assumed to vanish for $k \to 0$ and to diverge in the limit $k \to \infty$. We may write $R_k = k^n r_k$ with dimensionless $r_k$ typically depending on ratios of suitable differential operators $\mathcal{D}$ over the appropriate power of $k,\mathcal{D}/k^m$. (For gauge fields and scalars in four dimensions one has $n = m = 2$ and $\mathcal{D}$ will contain second covariant derivatives formed with the background field $\bar{g}$.) We require that $r_k$ vanishes fast for large $|\mathcal{D}/k^m|$, such that $\Delta S_k$ only affects the long distance modes. For small $|\mathcal{D}/k^m|$ we assume that $r_k$ approaches a constant. In the presence of the IR-cutoff $W$ and $\tilde \Gamma$ depend on the scale $k$. For $k \to 0$ the cutoff vanishes and $\tilde \Gamma_{k = 0}$ is the quantum effective action where all fluctuation contributions are included. For $k \to \infty$ the fluctuations are cut off and therefore not yet included in $\tilde\Gamma_k$. For an appropriate choice of $R_k$ the saddle point approximation becomes valid, such that $\tilde\Gamma_{k \to \infty} = S + S_\text{gf}+\Delta S_k$. In a slight modification of the usual treatment the cutoff \eqref{eqn:B4} acts directly on the fluctuations $g^\prime - \bar{g}$.

The flow of $W(L,\bar{g})$ with $k$ is given by the exact flow equation at fixed $L$ and $\bar{g}$,
\begin{alignedeqn}\label{eqn:B6}
	\partial_k W(L,\bar{g}) - \epsilon_k(\bar{g})
	&= -\frac{1}{2}\langle \partial_k\Delta S_k\rangle\\
	&= -\frac{1}{2} \Str\big\{\partial_k R_k(G^\prime + h \cdot h)\big\},
\end{alignedeqn}
with connected correlation function
\begin{equation}\label{eqn:B7}
	G_{ij}^\prime
	= \langle g^\prime_ig^\prime_j\rangle_c
	= \langle g^\prime_ig^\prime_j\rangle - \langle g^\prime_i\rangle\langle g^\prime_j\rangle,
\end{equation}
and
\begin{equation}\label{eqn:B7a}
	(h\cdot h)_{ij}
	= h_ih_j,
	\qquad
	h
	= g - \bar{g}
	= \langle g^\prime\rangle - \bar{g}.
\end{equation}
(The minus sign in the supertrace $\Str$ for fermions arises from the permutation of Grassmann variables.) Performing the Legendre transform \eqref{eqn:B2} at fixed $\bar{g}$ results in the flow of the effective action at fixed $g$ and $\bar{g}$
\begin{equation}\label{eqn:B8}
	\partial_k\tilde\Gamma (g,\bar{g})
	= \frac{1}{2} \Str\big\{\partial_k R (G^\prime + h \cdot h\big\} - \epsilon_k(\bar{g}).
\end{equation}
By virtue of the relation
\begin{equation}\label{eqn:B8A}
	\tilde\Gamma^{(2)} \, G^\prime
	= 1,
	\qquad
	\tilde\Gamma^{(2)ij}
	= \frac{\partial^2\tilde\Gamma(g,\bar{g})}{\partial g_i \partial g_j},
\end{equation}
the flow equation for $\tilde\Gamma(g,\bar{g})$ is closed in the two-field formalism.

Defining
\begin{equation}\label{eqn:B8B}
	\Gamma_k(g,\bar{g})
	= \tilde\Gamma_k(g,\bar{g}) - \frac{1}{2} h^\transp R_k(\bar{g})h,
\end{equation}
the second functional derivative obeys
\begin{equation}\label{eqn:B8C}
	\tilde\Gamma^{(2)}
	= \Gamma^{(2)} + R,
	\qquad
	G^\prime
	= (\Gamma^{(2)} + R)^{-1},
\end{equation}
and
\begin{equation}\label{eqn:B8D}
	\partial_k\Gamma
	= \frac{1}{2} \Str(\partial_k R G^\prime) - \epsilon_k.
\end{equation}
Eq. \eqref{eqn:B8D} has a one-loop form, with regulator $R_k$ appearing in the propagator $G^\prime$ according to \eqref{eqn:B8C}. For $k \to 0$ or $h \to 0$ the expressions for $\tilde\Gamma$ and $\Gamma$ coincide.

\section{Macroscopic field}
\label{Macroscopic field}

We will investigate a suitable choice of the macroscopic field $\bar{g}(g)$ and a suitable definition of a gauge-invariant effective action $\bar\Gamma(\bar{g})$ such that the propagator $G_P$ for the physical fluctuations can be expressed in terms of $\bar\Gamma$, typically involving the second functional derivative $\bar\Gamma^{(2)}$. This will produce the closed form \eqref{eqn:A} for the evolution equation of the gauge-invariant effective action. The exact flow equation \eqref{eqn:B8D} has been derived for an arbitrary background field $\bar{g}$ kept fixed. We want to translate this to a macroscopic field that is a function of $g, \bar{g} = \bar{g}_k(g)$. Here we have indicated that the relation between $\bar{g}$ and $g$ may depend on $k$.

Insertion of $\bar{g}(g)$ defines an effective action $\tilde\Gamma(g)$ that only depends on one argument
\begin{equation}\label{eqn:A1}
	\tilde\Gamma(g)
	= \tilde\Gamma\big (g,\bar{g}(g)\big).
\end{equation}
If the relation between $\bar{g}$ and $g$ is compatible with infinitesimal gauge transformations, i.e.
\begin{equation}\label{eqn:A2}
	\bar{g}(g + \delta_\xi g)
	= \bar{g} + \delta_\xi\bar{g},
	\qquad
	\delta_\xi\bar{g}
	= \frac{\partial \bar{g}}{\partial g}\delta_\xi g
	= \bar P\delta_\xi \bar{g},
\end{equation}
the effective action $\tilde\Gamma(g)$ is gauge invariant. This follows from the gauge invariance of $\tilde\Gamma(g,\bar{g})$ under the simultaneous transformation of $g$ and $\bar{g}$, with
\begin{equation}\label{eqn:A3}
	\delta_\xi\tilde\Gamma(g,\bar{g})
	= \frac{\partial\tilde\Gamma}{\partial g}\delta_\xi g + \frac{\partial \tilde\Gamma}{\partial\bar{g}}\delta_\xi\bar{g}
	= 0
\end{equation}
implying
\begin{equation}\label{eqn:A4}
	\delta_\xi\tilde\Gamma(g)
	= \frac{\partial \tilde\Gamma}{\partial g}(g,\bar{g})\delta_\xi g + \frac{\partial\tilde\Gamma}{\partial \bar{g}}\frac{\partial\bar{g}}{\partial g}
	\delta_\xi g
	= 0.
\end{equation}
This argument extends to a gauge-invariant effective action $\bar\Gamma$ which is related to $\tilde\Gamma$ by subtraction of a suitable gauge-invariant piece, similar to \eqref{eqn:B8B}. Replacing the argument $g$ by $\bar g$, i.e. inserting $g(\bar g)$, yields the gauge invariant action $\bar\Gamma(\bar g)$. 

A simple choice corresponding to the background field formalism would be the identification
\begin{equation}\label{eqn:A5}
	\bar{g}(g)
	= g.
\end{equation}
The resulting effective action $\tilde\Gamma(g)$ is gauge invariant. However, the first derivative of $\tilde\Gamma(g)$ produces no longer the source \cite{RW1},
\begin{alignedeqn}\label{eqn:A6}
	\frac{\partial\tilde\Gamma}{\partial g}
	= \frac{\partial\tilde\Gamma(g,\bar{g})}{\partial g} + \frac{\partial\tilde\Gamma(g,\bar{g})}{\partial \bar{g}}
	\frac{\partial\bar{g}}{\partial g}
	= L + \kappa,
\end{alignedeqn}
with
\begin{equation}\label{eqn:A7}
	\kappa
	= \frac{\partial \tilde\Gamma(g,\bar{g})}{\partial \bar{g}} \frac{\partial \bar g}{\partial g}.
\end{equation}
The matrix of second derivatives contains a generalized ``gauge fixing correction''
\begin{equation}\label{eqn:A8}
	Q^{ij}
	= \frac{\partial^2\tilde\Gamma(g,\bar{g})}{\partial g_i\partial g_j}-
	\frac{\partial^2\tilde\Gamma(g)}{\partial g_i\partial g_j}.
\end{equation}
This gauge fixing contribution appears in the exact flow equation \eqref{eqn:B8D} for $\Gamma(g) = \Gamma(g,g)$. Since $Q$ cannot be expressed in terms of $\Gamma(g)$ the flow of $\Gamma(g)$ is no longer given by a closed equation \cite{RW1}. It involves the generating functional $\tilde\Gamma(g,\bar{g})$ with two arguments, which has to be determined by some assumption or approximation. This is one of the main uncertainties in the present use of approximations to the flow equation.

The choice of $\bar{g}(g)$ is not unique, however. We will investigate a suitable choice such that the flow equation for the gauge-invariant effective action $\bar\Gamma(\bar{g})$ becomes closed. If $\bar\Gamma$ is gauge invariant its second derivative has zero eigenvalues. The propagator $G^\prime$ in the flow equation \eqref{eqn:B8D} can therefore not be given by the inverse of $\bar\Gamma^{(2)}$ for $k \to 0$. We will separate the ``physical fluctuations'' and ``gauge fluctuations'' by suitable projections. On the projected space of physical fluctuations $\bar\Gamma^{(2)}$ will become invertible and we will express $G_P$ by the inverse of $(\bar\Gamma^{(2)} + R)$ on this projected subspace, according to eq. \eqref{eqn:E}. The physical fluctuations will contribute the term $\pi_k$ in the flow equation \eqref{eqn:A}, while the gauge fluctuations are responsible for $\delta_k$.

\section{Projection on physical fluctuations}
\label{Projection on physical fluctuations}

The ``macroscopic fluctuations'' $h = g - \bar{g}$ can be split into ``physical fluctuations'' $f$ and ``gauge fluctuations'' $a$,
\begin{equation}\label{eqn:38A}
	h
	= g - \bar{g}
	= f + a.
\end{equation}
The gauge fluctuations $a$ obey
\begin{equation}\label{eqn:G4}
	P(\bar{g})a
	= 0,
	\qquad
	a
	= \big (1 - P(\bar{g})\big)h.
\end{equation}
On the other hand, $P(\bar{g})$ projects on the ``physical fluctuations'',
\begin{equation}\label{eqn:G5}
	f
	= P(\bar{g})h.
\end{equation}

For infinitesimal $h$ the gauge variation of $g$ at fixed $\bar{g}$ can be expressed as an inhomogeneous transformation of $h$,
\begin{equation}\label{eqn:38B}
	\hat \delta h
	= \bar P(\bar{g} + h)\delta_\xi(\bar{g} + h)
	= \bar P(\bar{g})\delta_\xi(\bar{g}) + \delta_h h.
\end{equation}
The inhomogeneous part only affects the gauge fluctuations $a$,
\begin{equation}\label{eqn:38C}
	\delta_\text{inh} a
	= \bar P(\bar{g})\delta_\xi(\bar{g}),
\end{equation}
while $f$ transforms as a tensor according to the homogeneous part $\delta_h f$. Infinitesimal gauge fluctuations $a$ can therefore be viewed as the result of an infinitesimal gauge transformation acting only on $g$.

Let us write $\Gamma(g,\bar{g})$, as defined in eq. \eqref{eqn:B8B}, in the form
\begin{equation}\label{eqn:G1}
	\Gamma(g,\bar{g})
	= \hat\Gamma(g,\bar{g}) + \Gamma_\text{gf}(g,\bar{g}).
\end{equation}
We assume that the ``gauge fixing term'' $\Gamma_\text{gf}$ is quadratic in the gauge fluctuations $a$
\begin{equation}\label{eqn:G3}
	\Gamma_\text{gf}(g,\bar{g})
	= \frac{1}{2\alpha}a^\transp \bar Q(\bar{g})a.
\end{equation}
We will take the limit $\alpha \to 0$ and assume that $\hat\Gamma$ remains finite in this limit. It will be important that no terms independent of $a$ or linear in $a$ diverge for $\alpha \to 0$. This selects a particular class of gauge fixing terms that are quadratic in $a$. Due to the divergence for $\alpha \to 0$ the term \eqref{eqn:G3} is the dominant contribution to $\Delta\Gamma(g,\bar{g})$ as defined by eq. \eqref{eqn:I1}.

For the second functional derivative the gauge fixing term contributes a term that diverges for $\alpha \to 0$,
\begin{equation}\label{eqn:G6}
	\Gamma^{(2)}_\text{gf}
	= \frac{1}{\alpha} \big (1 - P^\transp(\bar{g})\big)\bar Q(\bar{g})\big(1 - P(\bar{g})\big).
\end{equation}
The infrared cutoff is taken to contain a part $\bar{R}_k$ for the physical fluctuations as well as a cutoff for the gauge fluctuations $\sim R_{k,\text{gf}}$,
\begin{equation}\label{eqn:G7}
	R_k
	= \bar{R}_k(\bar{g}) + \frac{1}{\alpha} \big(1 - P^\transp(\bar{g})\big)
	R_{k,gf}(\bar{g})\big(1 - P(\bar{g})\big),
\end{equation}
such that the second functional derivative of $\tilde\Gamma$ obeys
\begin{equation}\label{eqn:G8}
	\tilde\Gamma^{(2)}
	= \hat \Gamma^{(2)} + \bar{R}_k + \frac{1}{\alpha} (1 - P^\transp)(\bar Q + R_\text{gf})(1 - P).
\end{equation}

We next decompose $\tilde\Gamma^{(2)}$ into four blocks corresponding to the different projections with $P$ or $\bar P = (1 - P)$ from left or right. The propagator $G$ can be decomposed similarly and one finds from eqs. \eqref{eqn:B8A}, \eqref{eqn:G8} for $\alpha \to 0$
\begin{equation}\label{eqn:G9}
	G
	= G_P + \alpha \, G_l,
	\qquad
	G_P
	= PGP^\transp.
\end{equation}
Here $G_P$ is the inverse of $\tilde\Gamma^{(2)}_P$ on the projected subspace
\begin{equation}\label{eqn:G10}
	\tilde\Gamma^{(2)}_PG_P
	= P^\transp,
	\qquad
	\tilde\Gamma^{(2)}_P
	= P^\transp(\hat \Gamma^{(2)} + \bar{R}) P.
\end{equation}
The piece $\alpha \, G_l$ vanishes for $\alpha \to 0$, and we observe that $\Gamma^{(2)}_\text{gf}$ and $R_\text{gf}$ do not contribute to $\tilde \Gamma^{(2)}_P$.

In the limit $\alpha \to 0$ the flow equation \eqref{eqn:B8D} consists of a part involving $G_P$ and a gauge contribution $\delta_k(\bar{g})$,
\begin{equation}\label{eqn:G11}
	k\partial_k\Gamma(g,\bar{g})
	= \frac{1}{2} \Str(k\partial_k\bar{R}G_P) + \delta_k - \epsilon_k,
\end{equation}
with
\begin{equation}\label{eqn:G12}
	\delta_k
	= \frac{1}{2} \tr\big\{k\partial_k R_\text{gf}(1 - P)
	(\bar Q + R_\text{gf})^{-1}(1 - P^\transp)\big\}.
\end{equation}
The gauge contribution arises from multiplication of $\alpha^{-1}\partial_k R_\text{gf}$ with the appropriate projection of $\alpha \, G_l$. It depends on $\bar{g}$ via $P,\bar Q$ and $R_\text{gf}$, but is does not involve $\hat \Gamma$. This defines the measure contribution $\delta_k$ in eq. \eqref{eqn:A}. We observe that for $\alpha \to 0$ only $\tilde \Gamma^{(2)}_P$ and the second term $\sim 1/\alpha$ in eq. \eqref{eqn:G8} enter in the flow equation \eqref{eqn:G11}. The parts in $\tilde\Gamma^{(2)} - \tilde\Gamma^{(2)}_P$ that do not diverge for $\alpha \to 0$ are projected out and do not influence the flow. We also observe that the leading part in $\Delta\Gamma$ is given by $\Gamma_\text{gf}$ in eq. \eqref{eqn:G3} for all $k$. All contributions to the flow are finite for $\alpha \to 0$ and therefore cannot change the divergent part $\sim 1/\alpha$. Furthermore, the r.h.s.\ of the flow equation contains no term that diverges for $\alpha \to 0$. As a consequence, one has $\partial_t (1/\alpha) = b_\alpha$ with finite $b_\alpha$, such that $\partial_t \alpha = -b_\alpha \, \alpha^2$ has a fixed point for $\alpha = 0$ \cite{CWEQ,CWGF}. Our gauge-fixing condition is not changed by the flow. All terms induced by the flow in the sector of gauge fluctuations are subleading and give vanishing contributions to $\pi_k$, $\delta_k$ and $\epsilon_k$ for $\alpha \to 0$.

We are finally interested in the flow of $\bar\Gamma(\bar{g})$, which is related to $\Gamma\big (g(\bar{g}),\bar{g}\big)$ by a suitable subtraction. We will choose $\bar{g}(g)$ such that $a(g) = 0$. Then $\bar{g}(g)$ is determined by specifying $f(\bar{g})$. Both $\delta_k$ and $\epsilon_k$ involve only $\bar{g}$. For the contribution $\pi_k$ of the physical fluctuations one needs the evaluation of the first term in eq. \eqref{eqn:G11} by inserting $g = \bar{g} + f(\bar{g})$. Furthermore, the flow at fixed $\bar{g}$ has to take into account that the relation between $g$ and $\bar{g}$ may depend on $k$. This flow equation for $\bar\Gamma(\bar{g})$ will be closed if we can find a suitable choice of $\bar{g}(g)$ such that $\hat \Gamma^{(2)}$ can be expressed in terms of $\bar\Gamma(\bar{g})$ and its functional derivatives. Then $G_P$ can be expressed in terms of $\bar\Gamma$ by solving eq. \eqref{eqn:G10}.

The particular form of the gauge fixing term \eqref{eqn:G3} is crucial for our construction. One may add other terms that do no diverge for $\alpha \to 0$, but there should be no term linear in $a$ that diverges for $\alpha \to 0$. Correspondingly, we employ a particular ``physical gauge fixing'' in the microscopic formulation \eqref{eqn:B1}
\begin{equation}\label{eqn:48A}
	S_\text{gf}(g^\prime,g)
	= \frac{1}{2\alpha}a^{\prime T} \bar Q(\bar{g})a^\prime,
\end{equation}
where
\begin{equation}\label{eqn:48B}
	a^\prime
	= \big (1 - P(\bar{g})\big)(g^\prime - \bar{g}).
\end{equation}
For $\alpha \to 0$ the saddle point approximation becomes exact in the sector of the gauge fluctuations and one infers the leading gauge fixing term in eqs. \eqref{eqn:G1}, \eqref{eqn:G3}.

For the example of Yang-Mills theories \eqref{eqn:48A} is realized by background gauge fixing in Landau gauge
\begin{equation}\label{eqn:48C}
	\Gamma_\text{gf}
	= \frac{1}{2\alpha}\int_x G^z G_z^*,
	\qquad
	G^z
	= \big (D^\mu(A^\prime_\mu - \bar A_\mu)\big)^z,
\end{equation}
with covariant derivative $D^\mu$ formed with the macroscopic field $\bar A_\mu$. With
\begin{equation}\label{eqn:48D}
	\bar Q
	= -D_\mu D^\nu,
	\qquad
	\mathcal{D}_S
	= -D^\rho D_\rho,
\end{equation}
and infrared cutoff for gauge fluctuations $R_\text{gf}(\bar Q)$, one obtains
\begin{alignedeqn}\label{eqn:48E}
	\delta_k
	&= \frac{1}{2} \tr \Big\{k\partial_k R_\text{gf}(\bar Q)\big (\bar Q + R_\text{gf}(\bar Q)\big)^{-1}\Big\}\\
	&= \frac{1}{2}\tr\Big\{k\partial_k R_\text{gf}(\mathcal{D}_S)\big(\mathcal{D}_S + R_\text{gf}(\mathcal{D}_S)\big)^{-1}\Big\}.
\end{alignedeqn}

For the Faddeev-Popov determinant,
\begin{alignedeqn}\label{eqn:48F}
	M
	&= \det\big(-D^\mu(\bar A)D_\mu(A^\prime)\big)\\
	&= \det \big (\mathcal{D}_S + iD^\mu(A^\prime_\mu - \bar A_\mu)\big)
\end{alignedeqn}
we may choose in eq. \eqref{eqn:B1A} the regularization
\begin{equation}\label{eqn:48G}
	E_k
	= \frac{\det\big(\mathcal{D}_S + R_\text{gf}(\mathcal{D}_S)\big)}{\Det(\mathcal{D}_S)}.
\end{equation}
This results in
\begin{equation}\label{eqn:48H}
	\epsilon_k
	= 2\delta_k.
\end{equation}
As advocated, $\mathcal{D}_S$ in $E_k$ and therefore $E_k$ and $\epsilon_k$ are fixed expressions of $\bar{g}$. We assume a similar type of regularization for quantum gravity. This contrasts with the alternative of introducing ghosts and regularizing the ghost propagator. In the formulation with ghosts the flow of the effective action for coupled gauge fields and ghosts has to be followed. The induced higher order ghost interactions enhance the complexity of the problem. We find it worthwhile to explore the possibility to regularize the Faddeev-Popov determinant by a fixed $\bar{g}$-dependent factor $E_k$.

\section{Choice of macroscopic field}
\label{Choice of macroscopic field}

We next use the freedom in the precise choice of the macroscopic field $\bar{g}(g)$ in order to obtain a closed flow equation for a suitably defined gauge-invariant effective action $\bar \Gamma(\bar{g})$. The main idea is to express the propagator for physical fluctuations $G_P$ in terms of the second functional derivative $\bar\Gamma^{(2)}$. This will determine the choice of $\bar{g}(g)$ and the precise definition of $\bar\Gamma(\bar{g})$.

Let us define
\begin{alignedeqn}\label{eqn:G14}
	\bar\Gamma(\bar{g})
	&= \Gamma\big(g = \bar{g} + f(\bar{g}),\bar{g}\big) - C(\bar{g})\\
	&= \hat\Gamma(g = \bar{g} + f(\bar{g}),\bar{g}) - C(\bar{g}),
\end{alignedeqn}
for some suitably chosen $f(\bar{g})$ and $C(\bar{g})$. This amounts to the choice $a(\bar g)=0$ or $\bar{g}(g) = g - f\big(\bar{g}(g)\big)$. The second derivative of $\bar\Gamma(\bar{g})$ becomes
\begin{alignedeqn}\label{eqn:G15}
	(\bar\Gamma^{(2)})^{ij}
	= \frac{\partial^2\bar\Gamma(\bar{g})}{\partial \bar{g}_i\partial \bar{g}_j}
	&= (\hat \Gamma^{(2)})^{ij}(g = \bar{g} + f,\bar{g})\\
	&\hphantom{{}=}+ B^{ij} - \frac{\partial^2 C}{\partial\bar{g}_i\partial\bar{g}_j}.
\end{alignedeqn}
The term $B^{ij}$ arises from derivatives of $\Gamma(g,\bar{g})$ with respect to $\bar{g}$, as well as from $\partial g_i/\partial\bar{g}_j = \delta^j_i + \partial f_i/\partial\bar{g}_j$,
\begin{alignedeqn}\label{eqn:G16}
	B^{ij}
	&= S^{ij} + \frac{\partial\hat \Gamma}{\partial g_m}\frac{\partial^2 f_m}{\partial\bar{g}_i\partial\bar{g}_j}\\
	&\hphantom{{}=}+ (\hat\Gamma^{(2)})^{mn}
	\left(\delta^i_m\frac{\partial f_n}{\partial\bar{g}_j} + \delta^j_n\frac{\partial f_m}{\partial \bar{g}_i} +
	\frac{\partial f_m}{\partial\bar{g}_i}\frac{\partial f_n}{\partial \bar{g}_j}\right).
\end{alignedeqn}
Here we define for the $\bar{g}$-dependence of $\hat \Gamma$ at fixed $g$
\begin{equation}\label{eqn:G17}
	\sigma^i(g,\bar{g})
	= \frac{\partial \hat \Gamma(g,\bar{g})}{\partial \bar{g}_i},
\end{equation}
and
\begin{equation}\label{eqn:G18}
	S^{ij}
	= \frac{\partial\sigma^i}{\partial\bar{g}_j} + \frac{\partial\sigma^j}{\partial g_i}
	 + \frac{\partial\sigma^i}{\partial g_j}
	 + \frac{\partial\sigma^j}{\partial g_m}
	\frac{\partial f_m}{\partial \bar{g}_i} + \frac{\partial\sigma^i}{\partial g_m}\frac{\partial f_m}{\partial \bar{g}_j}.
\end{equation}
All quantities are evaluated for $g = \bar{g} + f(\bar{g})$, $a(\bar{g}) = 0$.

There is a certain freedom in the choice of $f(\bar{g})$ and $C(\bar{g})$. The only requirement is that $f(\bar{g})$ transforms as a tensor and $C(\bar{g})$ is gauge invariant, and that $P^\transp BP$ can be expressed as a functional of $\bar\Gamma$. A possible simple choice determines $f(\bar{g})$ by a solution of the differential equation
\begin{equation}\label{eqn:G19}
	P_i{^l}\left(B^{ij} - \frac{\partial^2 C}{\partial \bar{g}_i\partial \bar{g}_j}\right)P_j{^k}
	= 0.
\end{equation}
This allows us to replace $\hat\Gamma^{(2)}$ by $\bar\Gamma^{(2)}$ in \eqref{eqn:G10} and therefore to close the flow equation. The solution of eq. \eqref{eqn:G19} depends on $C(\bar{g})$. Our aim is a simultaneous choice of $f(\bar{g})$ and $C(\bar{g})$ such that the flow equation remains simple.

The flow equation \eqref{eqn:G11} holds for fixed $g$ and $\bar{g}$. For the flow of $\bar\Gamma$ at fixed $\bar{g}$ we have to take into account that the solution $g(\bar{g})$ according to eq. \eqref{eqn:G19} will depend on $k$. With the definition \eqref{eqn:G14} one finds for the flow of $\bar\Gamma$ at fixed $\bar{g}$
\begin{alignedeqn}\label{eqn:G20}
	k\partial_k\bar\Gamma(\bar{g})
	&= \frac{1}{2} \Str(k\partial_k\bar{R} G_P) + \delta_k(\bar{g}) - \epsilon_k(\bar{g})\\
	&\hphantom{{}=}+ A_k(\bar{g}) - k\partial_k C(\bar{g}),
\end{alignedeqn}
with
\begin{equation}\label{eqn:G21}
	A_k(\bar{g})
	= \frac{\partial \Gamma(g,\bar{g})}{\partial g}k\partial_k f(\bar{g}).
\end{equation}
Here $\partial\Gamma(g,\bar{g})/\partial g$ has to be evaluated for fixed $\bar{g}$ at $g(\bar{g}) = \bar{g} + f(\bar{g})$, and we employ $\partial_k g_{|\bar{g}} = \partial_k f$. A possible simple choice employs
\begin{equation}\label{eqn:G22}
	k\partial_k C
	= A.
\end{equation}
Then the two last terms in eq. \eqref{eqn:G20} vanish. This realizes the flow equation \eqref{eqn:A}, \eqref{eqn:B}, \eqref{eqn:E}. The system of eqs. \eqref{eqn:G19}, \eqref{eqn:G22} determines both $f(\bar{g})$ and $C(\bar{g})$. It is rather complex. Fortunately, there is no need to solve this system in practice. It is sufficient to realize that a solution exists. Neither $\sigma$ nor $f$ or $C$ enter explicitly the proposed gauge-invariant flow equation. A choice of $\bar g(g)$ for which eqs. \eqref{eqn:G19} and \eqref{eqn:G22} hold for a suitable $C(\bar g)$ will be called ``optimal macroscopic field''.

\section{Optimal macroscopic field}
\label{Optimal macroscopic field}

We want to argue in favor of the existence of solutions to the system of equations \eqref{eqn:G19} and \eqref{eqn:G22}. For $\sigma^i = 0$ eqs. \eqref{eqn:G19} and \eqref{eqn:G22} have the simple solution
\begin{equation}\label{eqn:S1}
	f(\bar{g})
	= 0,
	\qquad
	C(\bar{g})
	= 0.
\end{equation}
This shows that non-zero $f$ and $C$ are related to the $\bar{g}$-dependence of $\hat \Gamma(g,\bar{g})$ at fixed $g$, and therefore to the $\bar{g}$-dependence of $W(L,\bar{g})$ at fixed $L$ in \eqref{eqn:B1}. The gauge fixing term does not contribute for $a = 0$. Its contribution to the second derivative \eqref{eqn:G6} would be projected out in eq. \eqref{eqn:G19}, and we have already defined $\hat\Gamma^{(2)}$ without a contribution from the gauge fixing term. The $\bar{g}$-dependence relevant for $\sigma^i$ can therefore only arise from $\Delta S_k(g^\prime,\bar{g})$ and
\begin{equation}\label{eqn:S2}
	S_{FP}(g^\prime,\bar{g})
	= -\ln M_k(g^\prime,\bar{g}).
\end{equation}

For a better understanding of $\sigma^i$ we need an expression for the $\bar{g}$-dependence of $\hat \Gamma$, which follows from the $\bar{g}$-dependence of $W$. The $\bar{g}$-dependence of $W(L,\bar{g})$ obeys
\begin{equation}\label{eqn:S3}
	\frac{\partial}{\partial\bar{g}_i}W(L,\bar{g})
	= -\langle \frac{\partial}{\partial \bar{g}_i}(\Delta S_k + S_{FP} + S_\text{gf})\rangle.
\end{equation}
With
\begin{equation}\label{eqn:S3-2}
	\frac{\partial\tilde\Gamma }{\partial \bar{g}_i}_{|g}
	= -\frac{\partial W}{\partial \bar{g}_i}_{|L}
\end{equation}
one obtains
\begin{align}\label{eqn:S4}
	\sigma^i
	&= \langle\frac{\partial }{\partial \bar{g}_i}(\Delta S_k + S_{FP})\rangle - \frac{1}{2} f_m\frac{\partial}{\partial\bar{g}_i}\bar{R}^{mn}f_n
	 + \bar{R}^{mi}f_i\notag\\
	&= \sigma_R^i + \sigma^i_{FP},\\
	\sigma_R^i
	&= \frac{1}{2}\Str\left\{\frac{\partial\bar{R}}{\partial\bar{g}_i}G_P\right\}.\notag
\end{align}
The regularized Faddeev-Popov determinant typically involves some operator $\tilde{\mathcal{D}}$, $M_k = \det (\tilde{\mathcal{D}})$, such that
\begin{equation}\label{eqn:S5}
	\sigma^i_{FP}
	= \left\langle\frac{\partial}{\partial\bar{g}_i}S_{FP}\right\rangle
	= -\left\langle \tr\left\{\left(\frac{\partial}{\partial\bar{g}_i}\tilde{\mathcal{D}}\right) \tilde{\mathcal{D}}^{-1}\right\}\right\rangle.
\end{equation}
For Yang-Mills theories and $k = 0$ the contribution of $\sigma_\text{FP}^i$ to the projected $B$ in \eqref{eqn:G19} may vanish for a suitable choice of the gauge fixing, but we have not yet investigated this issue.

The part $\sigma_R^i$ is proportional to $\partial \bar{R}/\partial \bar{g}_i$. It therefore vanishes for $k = 0$ where $\bar{R} = 0$. On the other hand, for large $k$ the cutoff function $\bar{R}$ approaches a $k$-dependent constant. In this limit $\sigma_R^i$ vanishes again. Thus $\sigma_R^i$ only plays a role in the range where typical differential operators are of a similar size as the appropriate power of $k$.

For small $\sigma^i$ we can solve the system of differential equations \eqref{eqn:G19} and \eqref{eqn:G22} iteratively. We split
\begin{equation}\label{eqn:X1}
	C
	= C_0 + C_1,
	\qquad
	C_0(\bar{g})
	= \frac{\partial\Gamma(g,\bar{g})}{\partial g_i}f_i,
\end{equation}
with $C_1$ obeying
\begin{equation}\label{eqn:X2}
	\gamma_k
	= \left(k\partial_k\frac{\partial\Gamma(g,\bar{g})}{\partial g_i}\right)f_i + k\partial_k C_1(\bar{g})
	= 0.
\end{equation}
For $B - C^{(2)}_0$ we observe that the second derivative of $C_0$ cancels the last term in eq. \eqref{eqn:G16} and the two last terms in eq. \eqref{eqn:G18}. We define
\begin{alignedeqn}\label{eqn:74A}
	\Delta B^{lk}_P
	&= P_i{^l}
	\biggl\{\tilde S^{ij} + \hat \Gamma^{(2)im}\frac{\partial f_m}{\partial \bar{g}_j} + \frac{1}{2}\hat\Gamma^{(2)mn}
	\frac{\partial f_m}{\partial \bar{g}_i}\frac{\partial f_m}{\partial \bar{g}_j}\\
	&\hphantom{{}=P_i{^l}
	\biggl\{}-\frac{1}{2}\frac{\partial^2 C_1}{\partial\bar{g}_i\partial \bar{g}_j} + (i\leftrightarrow j)\biggr\}P_j{^k},
\end{alignedeqn}
with
\begin{equation}\label{eqn:74B}
	\tilde S^{ij}
	= \frac{1}{2}\frac{\partial\sigma^i}{\partial\bar{g}_j} + \frac{\partial \sigma^i}{\partial g_j} - \frac{1}{2}\frac{\partial^2\sigma^i}{\partial\bar{g}_j\partial g_m}f_m,
\end{equation}
such that the condition \eqref{eqn:G19} reads $\Delta B_P = 0$.

In lowest order we consider $\sigma$ and $f$ as small quantities in which we linearize. One obtains $\Delta B_P = 0$ for
\begin{equation}\label{eqn:74C}
	\frac{\partial f_m}{\partial\bar{g}_j}
	= -(\hat G_P)_{mk}
	\left(\frac{\partial \sigma^k}{\partial g_j} + \frac{1}{2}\frac{\partial \sigma^k}{\partial \bar{g}^j} - \frac{1}{2}
	\frac{\partial^2C_1}{\partial\bar{g}^k\partial\bar{g}^j}\right),
\end{equation}
where
\begin{equation}\label{eqn:74D}
	\hat G_PP^\transp \bar \Gamma^{(2)}P
	= P.
\end{equation}
We may start with $C_1 = 0$, compute $f_m$ by solving the linear differential equation \eqref{eqn:74C} with suitable initial conditions, then determine $C_1$ for this solution from $\gamma_k = 0$, and iterate. From the linear solution higher order terms in $f$ and $\sigma$ can again be determined iteratively.

There seems to be no obstruction to find solutions to eq. \eqref{eqn:74C}. Typically, a particular solution will involve an initial condition that we may take as $f(\bar{g}_0) = 0$ for some suitably chosen configuration $\bar{g} = \bar{g}_0$. Linearization in $f$ is then expected to be valid for $\bar{g}$ in the vicinity of $\bar{g}_0$. These arguments do not constitute a proof that a solution of eqs. \eqref{eqn:G19}, \eqref{eqn:G22} exists for arbitrary $k$ and $\bar g$, even though this seems rather likely. A proof may be tried by starting at $k\to\infty$ with $C=0,f=0$, and computing the flow of these quantities by imposing the conditions \eqref{eqn:G19}, \eqref{eqn:G22}. If this solution breaks down the flow equation \eqref{eqn:A} is only an approximation, and we address the form of possible corrections in the discussion in sect. \ref{Discussion}. In a certain  sense we use the freedom in the choice of $\bar g(g)$ and $C(\bar g)$ in order to bring the form of an exact flow equation as close as possible to the form \eqref{eqn:A}. We stress again that the whole construction is only possible for a particular class of ``physical gauge fixings''.

\section{Quantum field equation}
\label{Quantum field equation}

For a given choice of $f(\bar{g})$ and $C(\bar{g})$ we can introduce a source $\bar J$ by
\begin{equation}\label{eqn:G23}
	\frac{\partial\bar\Gamma}{\partial\bar{g}}
	= \bar J.
\end{equation}
Gauge invariance of $\bar\Gamma$ implies
\begin{equation}\label{eqn:G24}
	P^\transp \bar J	
	= \bar J.
\end{equation}
This amounts to the usual conservation of $\bar J$. There is no need, however, that $\bar J^\transp$ equals precisely the projected ``microscopic source'' $J^\transp = L^\transp P$. The choice of $\bar{g}(g)$ determines the precise relation between $\bar{J}$ and $J$. The exact ``quantum field equation'' \eqref{eqn:G23} is the basis of the use of macroscopic field equations for gravity or electromagnetism, i.e. for general relativity and electrodynamics as ``classical field theories''. The appropriate ``quantum definition'' of the energy momentum tensor or the electromagnetic current is given by their coupling in the quantum effective action $\bar\Gamma$, as defined by eq. \eqref{eqn:G24}.

\section{Discussion}
\label{Discussion}

We propose a closed flow equation for the effective action of gauge theories that only involves one macroscopic gauge field - instead of separate fields for the expectation value of fluctuations and background field. Gauge invariance is maintained by this flow. This flow equation will be useful if $\bar\Gamma$ remains simple enough such that meaningful truncations can be devised and the ``initial value'' at large $k$ can be controlled. In particular, this concerns the locality properties of $\bar\Gamma$. The locality properties of $\bar\Gamma$ have to be found by practical computations for given models with gauge symmetry.

In particular, a gauge invariant effective action only involves the physical (``transverse'') fluctuations around a given solution of the field equations. There is no propagator for the gauge (``longitudinal'') fluctuations. Therefore no separate ``longitudinal gluon mass'' (or similar object in gravity) exists. A non-local mass term for gluons remains possible, however, induced by terms of the type $\sim F^{\mu\nu}f(-D^2)F_{\mu\nu}$, with $D^2$ a suitable differential operator acting on the physical fluctuations. Only a detailed computation can answer if our approach is useful to understand the infrared behavior of Yang-Mills theories. 

The simple form of the proposed gauge invariant flow equations hinges on the mere existence of a solution of eqs. \eqref{eqn:G19}, \eqref{eqn:G22}. (The precise form of the solution does not matter.) If not, the gauge invariant flow equation will contain a ``correction term'', $\zeta_k = \pi_k + \delta_k - \epsilon_k - \gamma_k$, with $\gamma_k$ defined by eq. \eqref{eqn:X2}, as well as a correction in the propagator equation \eqref{eqn:E},
\begin{equation}\label{eqn:Z1}
	\left(\bar\Gamma^{(2)}_P + \bar{R}_P + \Delta B_P\right) G_P	
	= P^{(T)},
\end{equation}
with $\Delta B_P$ given by eq. \eqref{eqn:74A}. If solutions for $f(\bar{g})$ and $C(\bar{g})$ with $\gamma_k = 0$, $\Delta B_P = 0$ exist, as suggested by our discussion above, the flow equation \eqref{eqn:A}, \eqref{eqn:B}, \eqref{eqn:E}, \eqref{XAB} is exact. If not, the proposed invariant flow equation can be still be used as an approximation, with errors $\sim f$. (The error can be minimized by the choice of optimal $f$ and $C$.)

Depending on the choice of $f(\bar{g})$ and $C(\bar{g})$ several versions of closed gauge-invariant flow equations for $\bar\Gamma(\bar{g})$ can be constructed. It is sufficient that $\Delta B_P$ and $\gamma_k$ can be expressed in terms of $\bar\Gamma(\bar{g})$. Besides gauge invariance, a simple structure and, in particular, a sufficiently local form of $\bar\Gamma$ are needed for devicing useful truncations. It is possible that a compromise with nonzero $\Delta B_P$ and $\gamma_k$ is advantageous for locality properties, even if solutions with $\Delta B_P = 0$, $\gamma_k = 0$ exist.

In practice, the non-local projections inherent in our approach are often not needed explicitly. The projections can be implemented by adding ``by hand'' a suitable ``physical gauge fixing''. In view of this, our approach argues in favor of the use of a particular ``physical gauge fixing'' that only acts on gauge fluctuations. For Yang-Mills theories this is realized by Landau gauge, e.g. \eqref{eqn:48C} with $\alpha \to 0$. For precise computations it may be advantageous from the point of view of locality properties of $\bar\Gamma$ to follow explicitly the flow of a ghost sector, computing the measure term $\epsilon_k$ from the contribution of the ghost fluctuations to the flow of the effective action, evaluated at nonzero gauge field $\bar A_\mu$ and vanishing ghost fields. Many computations of this type have been performed in the past in the background field formalism. They have neglected in practice the correction terms \cite{RW1} arising in the two-field formalism. We argue that this neglection can be justified. For $\alpha \to 0$ the gauge fixed flow equation, as employed so far, turns out to be identical to the projected flow equation \eqref{eqn:A}, \eqref{eqn:B}, \eqref{eqn:E}, \eqref{XAB}. Only the precise status of the macroscopic field $\bar A_\mu$ differs from the expectation value of the microscopic field if $f_\mu\neq 0$.

In quantum gravity the ``physical gauge fixing''
\begin{equation}\label{eqn:Z2}
	S_\text{gf}
	= \frac{1}{2\alpha}\int_x\sqrt{\bar{g}}\left(D^\mu h^\prime_{\mu\nu}\right)^2,
	\quad
	h^\prime_{\mu\nu}
	= g^\prime_{\mu\nu} - \bar{g}_{\mu\nu},
\end{equation}
involves the covariant derivative $D^\mu$ formed with the macroscopic metric $\bar{g}_{\mu\nu}$, and we take $\alpha \to 0$. This gauge fixing is purely quadratic in the gauge fluctuations. It has been advocated as the ``physical gauge fixing'' in ref. \cite{CWQC}, and used in practical computations in ref. \cite{Christiansen:2015rva}. Unfortunately, the algebraic complexity for this gauge is somewhat higher than for more popular gauges used in practical computations so far. These other gauges do not obey our criteria for the decoupling of gauge fluctuations. Corrections to simple truncations in the two-field formalism may therefore be substantial and are difficult to control \cite{MR,MRS,BEM}. We suggest to use the gauge fixing \eqref{eqn:Z2} and to employ the projected flow equation \eqref{eqn:A}, \eqref{eqn:B}, \eqref{eqn:E}. We hope that this helps to put the understanding of asymptotic safety in quantum gravity on a solid basis.

\medskip
\paragraph{Acknowledgment} The author would like to thank H.~Gies, J.~Pawlowski and M.~Reuter for useful comments and discussion. This work is supported by ERC-AdG \href{http://cordis.europa.eu/project/rcn/101262_en.html}{290623}.

\medskip
\paragraph{Note added} Since the first version of this work the proposed flow equation has been used in quantum gravity \cite{CWCC,CWIGR} and for Yang-Mills theories \cite{CWGI}. For the simple truncations employed there, no complications from the non-locality of projectors arose. A truncation for Yang-Mills theories based on an effective action $\sim F_{\mu\nu} \, F^{\mu\nu}$ produces the one-loop $\beta$-function for the running gauge coupling as well as $5/6$ of the two-loop coefficient. In this simple local truncation the flow of the gluon propagator contains no mass term.

\bibliography{References}

\end{document}